\documentstyle[preprint,tighten,aps]{revtex}
\begin{document}
\title{Interacting Boson Model plus broken-pairs description
of high-spin dipole bands}
\author{D. Vretenar$^1$, S. Brant$^1$, G. Bonsignori$^2$, L. Corradini$^2$,
C.M. Petrache$^{3,4}$}
\address{$^{1}$ Physics Department, Faculty of Science, University of
Zagreb, Croatia}
\address{$^{2}$Physics Department and INFN, University of Bologna, Italy} 
\address{$^{3}$Physics Department and INFN, University of Padova, Italy}
\address{$^{4}$ Institute of Physics and Nuclear Engineering, Bucharest,
Romania}
\date{\today}
\maketitle
\begin{abstract}
The Interacting Boson Model with broken-pairs has been extended to
include mixed proton-neutron configurations in the fermion
model space. The extended version of the model has been used to 
describe high-spin bands in the transitional nucleus $^{136}$Nd. Model
calculations reproduce ten bands of positive and negative parity states, 
including the two dipole high-spin structures based on the
$(\pi h_{11/2})^2$ $(\nu h_{11/2})^2$ configuration.
\end{abstract}
\vspace{1 cm}

\section{Introduction}
Models that are based on the interacting 
boson approximation \cite{IAC87} provide a consistent description
of nuclear structure phenomena in spherical, deformed and transitional nuclei.
Many extensions of the original Interacting Boson Model (IBM-1)
\cite{ArIa75} have been studied. In particular, models have
been constructed to describe the physics
of high-spin states in nuclei ($ 10\hbar \leq J \leq 30\hbar  $). 
In the formulation of these models one has to go beyond the boson
approximation and include selected 
non-collective fermion
degrees of freedom. By including part of the
original shell-model fermion space through successive breaking 
of correlated S and D pairs, the IBM is extended to describe
the structure of high-spin states. 
This extension of the model is especially relevant for transitional regions,
where single-particle excitations and vibrational collectivity are
dominant modes, and the traditional cranking approach to 
high-spin physics is not adequate.

The model that we present is an extension of our previous work
on the the physics of high-spin states in even-even and odd-even 
nuclei~\cite{IAC91,VRE93,VRE95}. 
The model is based on the IBM-1 \cite{IAC87}; 
the boson space consists of {\it s} and {\it d} bosons, with no distinction 
between protons and neutrons. To generate high-spin states,
the model allows one or two bosons to be destroyed and 
to form non-collective fermion pairs,
represented by two- and four-quasiparticle 
states which recouple to the boson core. 
High-spin states are described in terms of broken pairs.     
The model space for an even-even nucleus with $2N$ valence nucleons is
\begin{eqnarray}
\mid N~bosons~> & \oplus & \mid (N-1)bosons \otimes 1~broken~pair> \nonumber \\
& \oplus & \mid (N-2)bosons \otimes 2~broken~pairs> \oplus~~... \nonumber
\end{eqnarray}
The model Hamiltonian has four terms:
the IBM-1 boson Hamiltonian, the fermion Hamiltonian, the 
boson-fermion interaction, and a pair breaking interaction that 
mixes states with different number of fermions. In previous versions of 
the model for even-even nuclei, 
we could only separately consider neutron or proton excitations.
Mixed configurations were not included in the model space, i.e.
fermions in broken pairs had to be identical nucleons. 
For odd-A nuclei, we have extended the
IBFM~\cite{IAC79} to describe one- and three-fermion 
structures~\cite{VRE95}.
In that model mixed proton-neutron configurations have been 
included in the model bases. The two fermions in a
broken pair can be of the same type as the unpaired
fermion, resulting in a space with three identical fermions. 
If the fermions in the broken pair are different from the unpaired
fermion, the fermion basis contains two protons and one
neutron or vice versa. We have applied the model 
to the description of high-spin states in even-even and 
odd-even nuclei in the
Hg~\cite{IAC91,VRE93,VRE95}, Sr-Zr~\cite{CHO91,LIS93,CHI93,CAC96} and
Nd-Sm~\cite{PET97,DEA94,RAV96} regions. In several even-even 
nuclei high-spin bands have been observed, that could not
be described by model calculations. They were interpreted as
possible two proton - two neutron excitations, not included 
in our fermion model space.
In the present work we extend the model to include, 
in addition to two broken pairs of identical nucleons, 
the configurations with
two different broken pairs, one of protons and one of neutrons. 
We continue
our investigation of high-spin bands in the Nd-Sm region, 
and apply the model to the structure of positive 
and negative parity bands in $^{136}$Nd. This nucleus is 
especially interesting since, in addition to quadrupole bands,
a number of high-spin dipole bands have been observed. 
The occurrence of regular $\Delta J = 1$ sequences in 
nearly spherical or transitional nuclei is a relatively 
new phenomenon. It has attracted much interest, but this
is the first time that dipole bands in even-even nuclei
are described in the framework of a model that is based
on the interacting boson approximation. This means that 
we do not have to assume a geometrical picture for these
bands, and that they result from a consistent calculation
of the complete excitation spectrum, which includes also
the ground state band and two- and four-quasiparticle 
quadrupole bands. In Section II we present an outline 
of the model. The 
calculated excitation spectrum of $^{136}$Nd 
is compared with recent experimental data in Section III.
\section{The Interacting Boson Model plus broken pairs}
An even-even nucleus with $2N$ valence nucleons is described
as a system of $N$ interacting bosons.
The bosons represent collective fermion pair states
(correlated S and D pairs) that approximate the valence nucleon
pairs. In order to generate high angular momentum states, 
one or two bosons can be destroyed. They form noncollective pairs, 
represented by two- and four-quasiparticle states
that couple to the boson core. 
Fermions in broken pairs should in principle occupy all the valence
single-particle orbitals from which the bosons have been mapped.
However, for the description of high-spin bands 
close to the yrast line, the most important are the
unique parity high-j orbitals for which the Coriolis anti-pairing
effect is much more pronounced.
We define the model Hamiltonian
\begin{equation}
H=H_{B}+H_{F}+V_{BF}+V_{mix},
\end{equation}
where $H_B$ is the boson Hamiltonian of IBM-1~\cite{ArIa75},
and $H_F$ is the fermion Hamiltonian which 
contains single-fermion energies and 
fermion-fermion interactions. 
The interaction between the unpaired fermions
and the boson core contains the dynamical, exchange and
monopole interactions of the IBFM-1~\cite{IAC79}.
In order to describe high-spin dipole bands in 
$^{137}$Nd, in Ref.~\cite{PET97} we have modified
the quadrupole-quadrupole dynamical interaction 
\begin{equation}
V_{dyn}=\Gamma _{0}\sum_{j_{1}j_{2}}(u_{j_{1}}u_{j_{2}}-v_{j_{1}}v_{j_{2}})
\langle j_{1}\parallel Y_{2}\parallel j_{2}\rangle \times 
\left( [a^{\dagger}_{j_{1}} \times\tilde{a}_{j_{2}}]^{(2)}
\cdot Q^{B} \right),
\end{equation}
The standard boson quadrupole operator $Q^{B}$ 
\begin{equation}
Q^{B}=[s^{\dagger}\times\tilde{d}+d^{\dagger}\times\tilde{s}]^{(2)}
+\chi [d^{\dagger}\times\tilde{d}]^{(2)}
\end{equation}
has been extended by the higher order term
\begin{equation}
               \chi'\sum_{L_{1}L_{2}}\left[
                                     \left[
                                           d^{\dag} \times \tilde{d}
                                     \right]^{\left( L_{1} \right)}
                                     \times
                                     \left[
                                           d^{\dag} \times \tilde{d}
                                     \right]^{\left( L_{2} \right)}
                                     \right]^{\left( 2 \right)}\;.
\label{chi_prime}
\end{equation}
The boson-fermion interactions of IBFM were derived in the 
lowest seniority approximation~\cite{IAC79}. In the extension
of the model to high-spin states, most of the structures can
be described by the simplest form of the boson fermion interaction. 
However one should also expect polarization effects, as for 
example those that result in dipole bands, for which the 
boson-fermion interaction has to be extended by higher 
order terms. The term (\ref{chi_prime})
might arise as a nontrivial higher-order contribution in the
construction of the quadrupole moment from the generators of the U(6)
boson algebra.
For the exchange and monopole interactions we have retained
their standard forms.

The terms $H_B$, $H_F$ and $V_{BF}$ conserve the
number of bosons and the number of fermions separately. In our
model only the total number of nucleons is conserved, bosons
can be destroyed and fermion pairs created, and vice versa. In
the same order of approximation as for $V_{BF}$,
the pair breaking
interaction $V_{mix}$ which mixes states with different number of
fermions, conserving the total nucleon number only, reads
\begin{eqnarray}
V_{mix}&=&-U_{0}\left\{ \sum_{j_{1}j_{2}}u_{j_{1}}u_{j_{2}}(u_{j_{1}}v_{j_{2}}+
u_{j_{2}}v_{j_{1}})\langle j_{1}\parallel Y_{2}\parallel j_{2}\rangle ^{2}
\frac{1}{\sqrt{2j_{2}+1}}\left( [a^{\dagger}_{j_{2}}\times
a^{\dagger}_{j_{2}}]^{(0)}\cdot s\right)+\ h.c.\right\}  \nonumber\\
&&-U_{2}\left\{ \sum_{j_{1}j_{2}}(u_{j_{1}}v_{j_{2}}+
u_{j_{2}}v_{j_{1}})\langle j_{1}\parallel Y_{2}\parallel j_{2}\rangle
\left( [a^{\dagger}_{j_{1}}\times a^{\dagger}_{j_{2}}]^{(2)}
\cdot \tilde{d}\right) +\ h.c.\right\}.
\label{mixing}
\end{eqnarray}
This is the lowest order contribution to a pair-breaking interaction.
The first term represents the destruction of one $s$ boson and the
creation of a fermion pair, while in the second term a $d$ boson
is destroyed to create a pair of valence fermions.

If mixed proton-neutron configurations are included in the 
fermion model space, i.e. if broken pairs contain both protons and 
neutrons, the full model Hamiltonian reads
\begin{equation}
                H = H_{B} + H_{\nu F} + H_{\pi F} + V_{\nu BF}+
                 V_{\pi BF} + V_{\nu}^{mix} + V_{\pi}^{mix} + V_{\nu \pi},
\end{equation}
where the proton-neutron interaction term is defined 
\begin{eqnarray}
  V_{\nu \pi} & = & \sum_{\nu\nu'\pi\pi'} \sum_{J} (-)^{J}
                    h_{J}(\nu\nu'\pi\pi')
                    (u_{\nu}u_{\nu'}-v_{\nu}v_{\nu'})
                    (u_{\pi}u_{\pi'}-v_{\pi}v_{\pi'})
                                        \times \nonumber  \\
                                  &   & \times \left(
                                               \left[
                                 a^{\dag}_{\nu} \times \tilde{a}_{\nu'}
                                               \right]^{(J)} \cdot
                                               \left[
                                     a^{\dag}_{\pi} \times \tilde{a}_{\pi'}
                                               \right]^{(J)}
                                               \right)\;.
\end{eqnarray}
The two-body matrix elements of a residual proton-neutron interaction
define the coefficients $h_{J}(\nu\nu'\pi\pi')$:
\begin{eqnarray}
  <\nu\pi|V|\nu'\pi'> & = & \sum_{JM} h_{J}(\nu\nu'\pi'\pi)s_{\pi}s_{\nu'}
                          <j_{\nu}m_{\nu},\;j_{\nu'}-m_{\nu'}|JM>
                                               \times \nonumber  \\
                            &   & \times
                              <j_{\pi'}m_{\pi'},\;j_{\pi}-m_{\pi}|JM>\;,
\label{IDOH}
\end{eqnarray}
where $s_{\alpha}=(-)^{j_{a}-m_{\alpha}}$. 
In the present work we use the surface $\delta$-force (SDI) for the
residual interaction between unpaired fermions.
\section{The nucleus $^{136}_{~60}$Nd$_{76}$}
Nuclei in the A = 130-140 mass region are $\gamma$-soft and the 
polarizing effect of the aligned nucleons
induces changes in the nuclear shape.
In all even-even nuclei the alignment of both proton and
neutron pairs in the $h_{11/2}$ orbital generates 
low-lying 10$^+$ states, in many cases isomeric.
By coupling the proton and neutron pairs 
to collective core excitations, $\Delta J = 2$ decoupled bands
are generated with the 10$^+$ as band-head states.
Because of the different nature of the excitations (particles for 
proton, and holes for neutron configurations), the alignment of a pair of
$h_{11/2}$ protons drives the nucleus towards a collective shape 
opposite to that induced by the alignment of the neutron pair. 
One therefore expects to observe different 
coexisting structures at similar excitation energies. Employing
the IBM with broken pairs, we have described many
low-spin and high-spin properties of $\gamma$-soft
nuclei of this region (in the IBM language O(6) nuclei):
$^{137}$Nd ~\cite{PET97}, $^{138}$Nd~\cite{DEA94} and 
$^{139}$Sm~\cite{RAV96}.

In the present work we use the IBM with proton and neutron broken 
pairs to describe the excitation spectrum of $^{136}$Nd. In 
particular, we want to obtain a correct description 
of high-spin dipole bands which have been interpreted 
as two proton - two neutron structures. Experimental
data on high-spin structures in $^{136}$Nd have been 
recently reported in Refs.~\cite{PET96} and \cite{SUN96}.
In addition to two-proton and two-neutron quadrupole 
bands, as well as highly-deformed bands, five dipole
bands dominated by strong M1 transitions have been observed.
Theoretical calculations based on the projected shell model 
suggested quasiparticle configurations involving two protons and
two neutrons for four of these bands. The model interpretation 
for the fifth, highest lying band, was a configuration with four 
identical nucleons.
The experimental level scheme of positive-parity states
is displayed in Fig. \ref{figA}. The labels of bands are from 
Refs.~\cite{PET96} and \cite{SUN96}; 
in addition to the ground state band 
and the quasi $\gamma$-band, bands 3, 5, 7 and 8 
result from the alignment of two protons or two 
neutrons in the $h_{11/2}$ orbital, 
and the dipole bands 10 and 11
are based on four-quasiparticle configurations.

In $^{136}$Nd there are 6 neutron valence {\em holes} and 10 proton 
valence {\em particles}. 
The resulting boson number is N=8.
The set of parameters for the boson Hamiltonian is:
$\epsilon$=0.36, $C_0$=0.16, $C_2=-0.12$, $C_4=0.19$,
$V_2$=0.11 and $V_0=-0.15$ (all values in MeV).
The boson parameters have values similar to those 
used in the calculation of $^{137}$Nd, $^{138}$Nd 
and $^{139}$Sm.

In A $\approx$ 140 nuclei
the structure of positive parity high-spin states close to the 
yrast line is characterized by the alignment of both 
proton and neutron pairs in the $h_{11/2}$ orbital. 
For positive-parity states we have only included the 
proton and neutron $h_{11/2}$ orbitals in the fermion
model space. Additional single-nucleon states 
make the bases of the two broken-pairs prohibitively large. 
The single quasiparticle energies and occupation probabilities are obtained 
from a  BCS calculation using Kisslinger-Sorensen~\cite{KS63}
single-particle energies. For the proton $h_{11/2}$ orbital
the occupation probability is
$v^2_{\pi}(h_{11/2}) = 0.07$ and the
single-quasiparticle energy is $E_{\pi}(h_{11/2}) = 1.74$ MeV.
For neutrons $v^2_{\nu}(h_{11/2}) = 0.83$ and 
$E_{\nu}(h_{11/2}) = 1.13$ MeV. As we have shown 
in our previous calculations for 
$^{137}$Nd, $^{138}$Nd, and 
$^{139}$Sm, the
single-quasiparticle energy $E_{\nu}(h_{11/2})$$\approx$ 1.1 MeV is too low.
In order to reproduce the excitation energy of the
two lowest $10^+$ states, we have
renormalized the value of $E_{\nu}(h_{11/2})$ to 1.75 MeV, 
and of $E_{\pi}(h_{11/2})$ to 1.60 MeV.
If broken pairs are allowed to couple to all states of the 
boson core, the model bases are still too large.
In order to further reduce the size of the 
space with two-broken pairs, we prediagonalize the 
boson Hamiltonian, and 
then couple the fermion states to the lowest eigenvectors.
The collective structure of 
two- and four-quasiparticle high-spin bands is similar to 
that of the ground state 
band, but not identical, due to the smaller number of bosons.
The parameters of the fermion-boson interactions are determined from IBFM 
calculations of low-lying negative-parity states in $^{137}$Nd and 
neighboring odd-proton nuclei.
The parameters of the neutron dynamical fermion-boson interaction are
$\Gamma_0$=0.3 MeV, $\chi=-1$ and $\chi'=-0.2$, and for protons: 
$\Gamma_0$= 0.22 MeV, $\chi$=+1 and $\chi'$=+0.2.
The strength parameters of the exchange interactions are
$\Lambda_0^{\nu}=1.0$ and
$\Lambda_0^{\pi}=1.5$ for neutrons and protons, respectively.
The strength parameter of the pair-breaking interaction is $U_0=0$ and
$U_2=0.2$ MeV, both for protons and neutrons in broken pairs.
The residual interaction between unpaired fermions is a surface 
$\delta$-force with strength parameters
$ V_{\nu\nu} = -0.1 $ MeV, $ V_{\pi\pi} = -0.1$ MeV and
$ V_{\nu\pi} = -0.9 $ MeV for neutron-neutron, proton-proton and 
neutron-proton configurations, respectively.

In Fig. \ref{figB} we display the calculated spectrum of positive-parity 
states. Only few lowest states of each angular momentum are included. 
According to the structure of wave functions, states 
are classified in bands labeled in such a way that a direct
comparison can be made with their experimental counterparts.
The calculated positive-parity structures 
3, 5, 7, 8, 10 and 11, as well as the ground state band and the 
quasi $\gamma$-band, have to be compared with the experimental bands 
of Fig. \ref{figA}. The collective ground state band is 
the yrast band up to angular momentum $J = 8^+$. The calculation
reproduces the experimental positions of states of the ground state
band, as well as the quasi $\gamma$-band: $2^+_2$, $3^+_1$, and
$5^+_1$. The bands 3, 5, 7, and the sequence 
of levels $14^+$, $16^+$ and $18^+$ on the left of band 7, 
result from the alignment
of a pair of protons in the $h_{11/2}$ orbital. 
The main components 
in the wave functions of states of bands 3 and 5 are:
$|(\pi h_{11/2})^2 J_F = 10, I_B; J = J_F + I_B (-1)>$, where
($-1$) refers the odd-spin band, and $|I_B>$ denotes the lowest
collective state of the boson core. In band 7 the two protons 
are predominantly coupled to 
$J_F = 8$, and for $J \geq 14$ there is 
also a large contribution from  components
with $J_F = 10$. Band
8 corresponds to two aligned $h_{11/2}$ neutrons 
with $J_F = 10$, coupled to the boson core. We notice that 
the model calculation reproduces in detail the quadrupole 
collective two-proton and two-neutron bands. 
Finally, the two dipole bands 10 and 11 correspond to four-quasiparticle
states (two broken pairs, one proton and one neutron). 
The states 
$[(\pi h_{11/2})^2 J_F = 10]$$[(\nu h_{11/2})^2 J_F = 10]$ 
are coupled to the ground state band of the core. 
Bands 10 and 11 correspond to the two lowest
calculated states for each angular momentum of 
the configuration $(\pi h_{11/2})^2$ $(\nu h_{11/2})^2$. The 
calculated structures are in very good
agreement with the experimental bands.

The occurrence of regular dipole bands ($\Delta J=1$) in nearly 
spherical and transitional nuclei presents an interesting phenomenon. 
In the semiclassical picture of the cranked shell model,
$\Delta J=1$ high-spin bands have been described 
as TAC (Tilted Axis Cranking) solutions~\cite{fra93,fra97}. 
In our model such $\Delta J=1$ structures result from the 
fermion-boson interaction of unpaired fermions with the core, as well
as from the residual proton-neutron interaction between unpaired 
fermions in unique-parity high-j orbitals. In order to obtain
the correct energy spacings for bands 10 and 11, it was necessary
to include the additional term~(\ref{chi_prime}) in the 
boson quadrupole operator. We have also found that 
the proton-neutron $\delta$-interaction
plays a crucial    
role in the excitation spectrum of these bands.  
This is illustrated 
in Fig.~\ref{figC}, where we display three steps in the 
construction of band 11, and compare them with the 
experimental band. In the left column we plot results
of model calculation obtained with parameters
described above, except that the dynamical fermion-boson 
interactions for protons and neutrons 
do not contain the term~(\ref{chi_prime}),
and the proton-neutron $\delta$-interaction 
is not included.
The structure essentially consists of two quadrupole
bands, with almost degenerate doublets of odd and even 
spin states. This is a structure one normally expects
to find in transitional $\gamma$-soft nuclei. The 
quasi-degeneracy of the doublets is removed in 
step two, where the term (\ref{chi_prime}) has
been included in the dynamical interactions for 
protons and neutrons. Finally, the correct excitation 
energies relative to the band-head 16$^+$ 
are obtained in step three with the inclusion
of the proton-neutron residual interaction.

It should be noted that
bands 10 and 11 have been recently described in the 
framework of the projected shell model~\cite{SUN96}. This model 
uses a fixed deformation and provides a good description 
for the near yrast spectra of well deformed nuclei 
with stable shape. In order 
to obtain bands of dipole character, an axially symmetric shape 
with deformation $\epsilon_2 = 0.20$
had to be assumed, which in turn made energetically more favorable
one of the neutrons to occupy the $\nu f_{7/2}$ orbital. Therefrom
the configuration $(\pi h_{11/2})^2$ $\nu f_{7/2}$ $\nu  h_{11/2}$
was assigned to bands 10 and 11.
The dipole character of the bands results from the coupling 
of the neutron hole in $h_{11/2}$ and the neutron particle
in $f_{7/2}$. This interpretation does not agree with the 
results of our calculation. When included in our model   
space, the structures based on the $\nu f_{7/2}$ orbital
are found high above the yrast. Since $(\pi h_{11/2})^2 J_F =10$
and $(\nu h_{11/2})^2 J_F =10$ are the lowest two-fermion
bands, it is hard to see how could it be that four-fermion bands 
based on these configurations are not observed in the 
experiment. A larger deformation induced by polarization
effects could place these configurations higher
above the yrast. Such deformation however, should then
result from the dynamics, and not as an additional parameter.  
We notice that a very similar
dipole band, at approximately the same excitation energy, 
was found also in $^{138}$Nd~\cite{DEA94}. This nucleus is closer
to the $N=82$ shell, and therefore deformations 
induced by polarization effects should be even smaller.
On the other hand, using a unique set of 
parameters, we have been able to reproduce the complete 
experimental spectrum of positive parity states, from the 
ground state band 
to band 11, up to angular momentum 29 $\hbar$ 
at more than 13 MeV excitation energy. 

The structure of negative parity bands is displayed in 
Figs.~\ref{figD} and \ref{figE}. The labels are those 
of Ref.~\cite{SUN96}. As in the case of $^{138}$Nd~\cite{DEA94},
we assume that bands 1 and 2 correspond to a two-neutron
configuration. For the calculation of negative parity bands 
the fermion part of the model space includes one and two 
neutron pairs. In addition to $h_{11/2}$, 
the neutron orbitals
$s_{1/2}$ and  $d_{3/2}$ are included in the model space.
The quasiparticle energies and occupation probabilities:
$E_{\nu}(d_{3/2}) = 0.91$ MeV, $v^2_{\nu}(d_{3/2}) = 0.32$
and $E_{\nu}(s_{1/2}) = 0.88$ MeV, $v^2_{\nu}(s_{1/2}) = 0.64$.
The neutron parameters are the same as in the calculation
of positive parity states, except that 
$\Lambda_0^{\nu}=0$ and
$ V_{\nu\nu} = -0.54 $ MeV. These are effective interactions, 
and therefore the strength parameters depend on the fermion
model space. However, the precise values are not that important
for the structure of bands. In this particular case, the 
absence of the exchange interaction affects the moments 
of inertia of the bands, and the strength of the 
$\delta$-interaction is determined from the energy 
splittings between even and odd spin states. The lowest
calculated states are compared with the experimental bands 
1 and 2 in Fig.~\ref{figD}. The theoretical bands are in
good agreement with experimental data. The bands are based
on the $(\nu d_{3/2}, \nu h_{11/2})$ configuration, coupled
to the collective ground state band. In the lower part
of the spectrum there are also significant contributions
from components based on the configuration
$(\nu s_{1/2}, \nu h_{11/2})$, but generally states based
on this configuration are higher in energy. We have also 
investigated the possibility that the lowest negative
parity bands are based on two-proton configurations
in orbitals $h_{11/2}$, $d_{5/2}$ and $g_{7/2}$
as suggested in Ref.~\cite{PET96}, 
but the calculation did not reproduce the experimental 
excitation energies. The lowest negative parity
four-neutron states that result from our calculation 
are compared with band 12 in Fig.~\ref{figE}. The calculated
band is based on the configuration $\nu d_{3/2}(\nu h_{11/2})^3$
coupled to the boson core,
and reproduces the experimental spectrum. Of course, 
the possible assignment for band 12 could also be 
$(\nu d_{3/2}, \nu h_{11/2}) (\pi h_{11/2})^2$,
as suggested in Ref.~\cite{SUN96}. 
However, we were not able to verify this
assignment, since the fermion model space 
for negative parity states with proton and 
neutron pairs is simply too large for any 
practical calculation. The experimental spectrum 
contains another negative parity structure (band 9 of
Ref.~\cite{SUN96}), based on a state $9^-$ at 3182 keV.
The density of calculated negative parity states in this
region is very high. States have strongly mixed wave functions,
and we could not find a corresponding theoretical sequence.

In conclusion, we have used an extension of the 
Interacting Boson Model with broken pairs to describe 
high-spin dipole bands in transitional nuclei.
The model contains mixed proton-neutron configurations, 
i.e. proton and neutron broken pairs.
Compared with models based on the cranking approximation,
the present approach provides several advantages. 
High-spin structures can be
described not only in well deformed,
but also in transitional and spherical nuclei. A single set of parameters
and a well defined Hamiltonian are used to calculate 
collective bands of the core, high-spin two- and
four-quasiparticle bands. Polarization effects directly
result from the model fermion-boson interactions.
All calculations are performed in the laboratory frame, 
and therefore the results can be
directly compared with experimental data.
For the transitional nucleus $^{136}$Nd, model 
calculation reproduce the complete experimental excitation 
spectrum of positive and negative parity states. In particular,
we have been able to obtain a correct description of the 
two high-spin $(\pi h_{11/2})^2$ $(\nu h_{11/2})^2$ dipole bands.


%
%
\begin{figure}
\caption{ Experimental excitation spectrum of positive-parity
states in $^{136}$Nd.}
\label{figA}
\end{figure}

\begin{figure}
\caption{Results of IBM plus broken pairs calculation
for positive-parity bands in $^{136}$Nd.}
\label{figB}
\end{figure}

\begin{figure}
\caption{Steps in the construction of band 11 
compared with the experimental excitation spectrum.  
See text for details.}
\label{figC}
\end{figure}

\begin{figure}
\caption{Negative-parity bands 1 and 2 compared with results
of the IBM plus broken pair calculation.
Only states with possible experimental
counterpart are shown.}
\label{figD}
\end{figure}

\begin{figure}
\caption{Results of model calculation compared 
with the experimental negative-parity band 12.} 
\label{figE}
\end{figure}

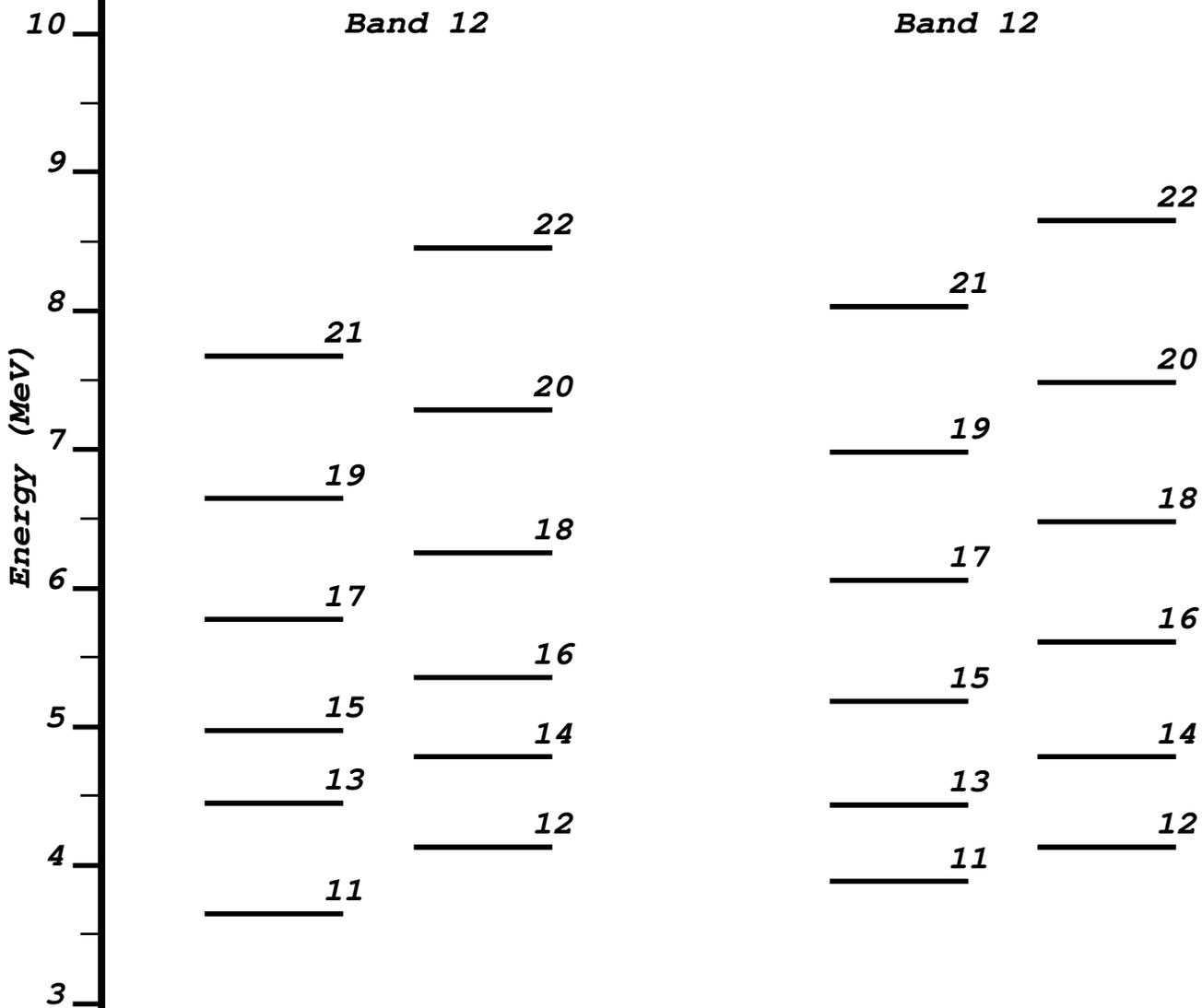
\end{document}